**Rick Szostak**
**University of Alberta**

**Richard P. Smiraglia**
**University of Wisconsin-Milwaukee**

**Andrea Scharnhorst**
**Data Archiving & Networked Services (DANS)**

**Aida Slavic**
**UDC Consortium**

**Daniel Martínez-Ávila**
**Universidad Carlos III de Madrid**

**Tobias Renwick**
**University of Alberta**

# Chapter 2
# Classifications as Linked Open Data
**Challenges and Opportunities**[‡]

**Abstract**

Linked Data (LD) as a web-based technology enables in principle the seamless, machine-supported integration, interplay and augmentation of all kinds of knowledge, into what has been labeled a huge knowledge graph. Despite decades of web technology and, more recently, the LD approach, the task to fully exploit these new technologies in the public domain is only commencing. One specific challenge is to transfer techniques developed pre-web to order our knowledge into the realm of Linked Open Data (LOD). This paper illustrates two different models in which a general analytico-synthetic classification can be published and made available as LD. In both cases, an LD solution deals with the intricacies of a pre-coordinated indexing language. The Universal Decimal Classification (UDC) approach illustrates a more complex solution driven by the practical requirements that the LD model is expected to fulfill in the bibliographic domain, and within the constraints of copyright protection. The Basic Concepts Classification (BCC) is a new classification with a novel approach to classification structure and syntax for which LD is an important vehicle for increasing the scheme's visibility and usability. The report on these two cases illustrate some of the challenges of the representation of knowledge organization systems as LD and the possibilities that analytico-synthetic and interdisciplinary or phenomenon-based systems present for the representation of knowledge using LD.

---

[‡] Reprinted with minor editorial emendations by permission from *Knowledge Organization at the Interface: Proceedings of the Sixteenth International ISKO Conference, 2020 Aalborg, Denmark*, ed. Marianne Lykke, Tanja Svarre, Mette Skov and Daniel Martínez-Ávila. Advances in Knowledge Organization 17. Baden-Baden: Ergon Verlag, 436-45.

## 1.0 Introduction

There is much excitement about the introduction of formal systems of knowledge organization (KO) into the infrastructure of the Linked Data (LD) and especially Linked Open Data (LOD) cloud. Expectations are grounded in the fact that LD connect phenomena with shared (controlled) vocabularies. In theory, meaningful links from specific points in the cloud-based knowledge graph to normalized concepts in formal classifications can help to strengthen a shared conceptual infrastructure—not simply meaningful semantics but also effective syndetic routing among concepts. This objective was the core research question of the "Digging Into the Knowledge Graph" research project.[1]

Szostak et al. (2018, 527-28) pointed out how three major challenges comprised sorting concepts, translating across domains and publishing knowledge organization systems (KOSs) as LOD. The Universal Decimal Classification (UDC) and the Basic Concepts Classification (BCC) —one disciplinary and the other phenomenon-based—were chosen as case studies to explore what problems might emerge along the journey of making KOSs available as LOD. As Siebes at al. (2019) detailed, the process of moving into the realm of LD is composed of stages of conceptual and technological explorations and decisions. Under the former fall questions such as what information to make available in a machine-readable form, to which extent existing vocabularies should be re-used, and whether or how to already enrich your LD prior to publication. Under the latter we find questions such as which web domain to use, how to design the URL's, but also how to guarantee stability over time and how to document provenance during possible editions or versions of the LD publication.

Special challenges arise from the formal representation of KOSs as LD which are at once semantic and logistical. Semantic issues arise due to terminological diversity in the unorchestrated, self-organized nature of the LD cloud itself. The job of linkage from specifically well-defined points in a classification to a potential of semantic relations in the cloud is a non-trivial research task. Methodologically, and when dealing with Linked Open data (LOD), different routes for interlinking exist: point-to-point explorations in the process of publishing a resource as LOD (Siebes et al. 2019); inspection of LOD clusters as literary warrant (Martínez-Ávila et al. 2018, 10); and translation between knowledge domains (Eito-Brun 2018; Marcondes 2018).

For KOSs that come with an extended legacy (a long history of well curated editions), such as the UDC, the choice of the appropriate namespace is non-trivial. We report approaches taken to publish the UDC and BCC as LOD enabling seamless integration into the cloud. Problems tackled in the process encompass data modelling, design of applied web technology (e.g., URI design), versioning (instantiating), licensing, extending KOSs published as LOD, and other possibilities to disseminate, exploit and enhance KOSs. Publication of a KOS as LD is not trivial; rather, it requires a whole process of which many parts need to be accomplished first off-line.

The task of translating a KOS into LOD is challenging in many ways. In the aforementioned conceptual stage, selectivity is one aspect seldom discussed. The first task is not to transfer the whole of the KOS to the new (Resource Description Framework or RDF) data model, but to choose those parts of the KOS that are most importantly available in a machine-readable LD format. In this process, the use of already existing vocabularies is reco mmended to express selected features from the KOS in the new data model. This translation may allow, or at least facilitate, a translator to translate KOS terminology into items already mapped into LOD schemas. A second task is that of mapping connections from one RDF schema to another. Both tasks are far from being a mechanical mapping process, but rather require research as exemplified further below.

More particularly, this paper illustrates two different models in which a general analytico-synthetic classification can be published and made available as LD. In both cases, an LD solution deals with the intricacies of a pre-coordinated indexing language. The UDC approach illustrates a more complex solution driven by the practical requirements that the classification LD are expected to fulfil in the bibliographic domain, and within the constraints of copyright protection. The BCC, interdisciplinary in nature, is a new classification with a novel approach to classification structure and syntax for which LD are an important vehicle for increasing the scheme's visibility and usability.

## 2.0 The BCC linked data publishing model

The Basic Concepts Classification (BCC)[2] was created by Rick Szostak for the purpose of providing structured direct access by phenomenon to documents (and the ideas expressed in them). The BCC grew by the addition of schedules of mostly verb-like relators and adjectival/adverbial properties added to the original schedule of phenomena. Documents (objects, ideas, concepts) can be expressed with combinations of phenomena, relators and properties, either in symbolic notation (classified form) or in natural-language sentence style.

The primary difficulty in mapping a universal (i.e. a general) KOS such as the BCC to LOD is that the BCC is intended to be able to classify almost anything (see Szostak 2019 for an overview of the BCC). The LOD cloud is also universal in extent, but achieves this universality with millions of distinct terms of varying degrees of specificity. A perfect mapping of the BCC to LOD would be able to encompass the entire cloud, but only by expanding BCC classes to such an extent that they would cease to be useful for classificatory purposes. The translator is left with the task of selecting points in the LOD cloud that hopefully encompass as much related information as possible.

The first task of the translator is to understand the relations, overlap and accepted usage among the current LOD cloud schemas. The initial impression on the translator is a bewildering array of options, some new, growing and maintained (e.g., Wikidata, DBPedia, OWL, SKOS, FOAF) and others suffering from disuse, abandonment or deprecation (e.g., Freebase). This array of options is a strength of LOD, for anyone can say anything about any topic (this is the so-called AAA rule that governs the semantic web), but for the translator it is very daunting to try to figure out whether someone else is trying to say the same thing as your KOS.

Within the BCC there are essentially nouns (phenomena) and verbs (relators). There are also adjective-like Properties that can be treated in much the same way as nouns. Phenomena are significantly simpler to map as the translator needs to choose a sufficiently large schema and map terms in the BCC directly to those matching entities. As an example, we have mapped the phenomenon of "art" (http://purl.org/basic/a-art), using the relation of "sameAs" from OWL (http://www.w3.org/2002/07/owl/sameAs), to the DBPedia entry on Art (http://dbpedia.org/resource/Art). This is a reasonable mapping, and it implies that anything anyone has classified as art using DBPedia is also classified as art in the BCC. Note that this implies to a graph query engine that the terms of the BCC and DBPedia are identical and can be merged. This property may not be ideal. In our example, within DBPedia a movie is not art, but rather it is a subclass of work (also it is identified as equivalent to schema.org "movie" where it is a creative work). In the BCC, film is indeed a subclass of art (http://purl.org/basic/ar2-film) meaning that the classification is now disjoint and films classified as art in the BCC will map to DBPedia incorrectly (at least according to DBPedia's definition). This is a general problem, for a controlled vocabulary such as the BCC is generally of greater breadth (that is, each term has a broader meaning) than an uncontrolled vocabulary such as the LOD Cloud.

One of the first points that the translator needs to comprehend is that it is possible to indicate the cardinality of relationships within LOD. For example, within SKOS there are classifications for broader and narrower, where the former implies that the object of the triple is broader than the subject, and narrower implies the inverse. For BCC "art" one might be tempted to say that "DBPedia:art SKOS:broader BCC:art", which indicates that everything DBPedia considers art is art in the BCC, but not everything the BCC considers art is in DBpedia art. The downside of using broader and narrower is that the mapping of the reciprocal is ambiguous (there is no way to know whether a BCC-Art object should be DBPedia Art). Further, there are likely examples mapped to DBPedia-Art that are not in BCC-Art. The true cardinality of the relation is that there is a significant amount of overlap between DBPedia-Art and BCC-Art, and therefore we reasonably consider them the same, given our goals. That is, we allow some small degree of inaccuracy in translation in order to indicate a broad overlap in meaning.

For the translation of relators the task is compounded as relators are used in the BCC to tie phenomena to one another, but in a more lexical way than LOD. In the BCC, relators can be used in conjunction with phenomena to add specificity to the classification. In terms of LOD, the word "visual" could be represented as "by pictures (/T7p)," where "by" is a relator and "pictures" is a phenomenon, but the idea represented by the two terms is smaller in scope than either term together. The translator may want to consider the effect of mapping the word "by" to any other definition, as while they may appear to be similar, what this implies is that for an object already mapped in LOD to be mapped to BCC it would have to link to both items in some way, which is unlikely. Again, here we were faced with a decision of imprecision and decided to create an independent classification for the relator "by."

The goal of translating a KOS to LOD is not to achieve perfection, but rather to cast a broad enough net so that the first iteration of the KOS can bring in terms that are related closely enough to its topics to test whether the KOS is capable of their classification. To this end, we begin with accepting the imperfect and hoping that it allows for iterative improvement.

## 3.0 The UDC linked data publishing model
The UDC has been one of the most widely used KOSs in the bibliographic domain for over a century. It is often used in conjunction with and complementary to thesauri, subject heading systems, and special classifications. During its lifetime, the classification has undergone many changes and has been made available in many languages and versions. The current UDC data standard, the UDC Master Reference File (UDC MRF) has had over twenty updates released since 1993, with 50% of the current 72,000 sets of classes having been added or changed. The UDC data also include 12,000 cancelled (deprecated) classes that redirect to new classes. The scheme is currently owned, maintained, and developed by an international consortium of publishers, on a self-funding and non-profit basis.

The UDC scheme organizes concepts and subjects within traditional forms of knowledge (disciplines) allowing concepts and classes of concepts to be freely combined both within and between subject fields to express any level of complexity that information resources present. When both classification schemes and bibliographic metadata are published as linked data and are connected, they form a complex and dynamic knowledge space that shows the ways we create, interact with or utilize information. Classmarks stored in millions of bibliographic records hold valuable information about the contents of these collections. Once UDC classmarks are linked back to the classification scheme from which they originated, it is possible to capture their meaning and establish further meaningful associations within and among collections (cf. Slavic 2017). These connections made through linked data can help to:

- enrich bibliographic data to support information discovery by increasing subject access points using UDC terminology, by enabling semantic expansion (broadening); and by improving precision through contextualization;
- improve systematic presentation, grouping, and visualization of resources and collections (linear or multidimensional) to facilitate browsing and serendipitous discovery of information;
- link the classification to other KOSs to enable cross-collection information discovery; and,
- validate and update local classification data and local authority files or bypass local and obsolete classification data in information exchange.

Apart from many practical aspects of interest, UDC LD development represents a good testbed for further research especially through its interaction with other KOSs. As an example of an analytico-synthetic and faceted scheme, it provides a case study for managing the alignment between the simple codes that appear in the scheme and the complex classmarks generated through document indexing that contain unlimited numbers of combinations of UDC classmarks.

## 3.1 Challenges and solutions

While longevity and widespread use represent strong arguments for sharing the UDC as LD, this also requires more responsibilities and presents further difficulties. In 2011, an extract from UDC of 2,600 classes was published as LOD in SKOS format. This experiment proved to be a valuable experience. As more and more library catalogs became available as linked data, we learned about the magnitude of the incompatibilities between UDC classmarks in bibliographic records and the UDC standard data.

Library linked data (LLD) clouds that were observed contained specific and complex UDC classmarks that could only be resolved through the access to the complete UDC content. However, the main cause of mismatch between UDC namespace and LLD is in the fact that libraries continue to use deprecated UDC codes. Thus, it became clear that a UDC namespace has to include not only the complete content of the UDC MRF, but also a significant collection of historical data and concordances between cancelled and new classes. Needless to say, the UDC LD used from 2011-2019 indicated that programs utilizing the UDC namespace (or those creating them) have little awareness of the UDC data structure, semantics, syntax, provenance, versioning, and changes and might not be able to process and select UDC data accurately or make good use of them.

In order to serve its purpose in a bibliographic domain, the UDC namespace has to provide a robust solution for the linking and semantic alignment between classmarks in bibliographic records and those in the UDC LD cloud. This has to be achieved irrespectively of the fact that the classmark strings in library data include combinations of simple UDC codes or that some may be deprecated or generated through wrong local practice. In order to achieve this, important changes had to be made to the ways and format in which UDC LD is published. This included the change of the URI format and the change of the RDF schema, but most importantly, instead of a UDC LD dump we opted for a more complex UDC look-up service.

The main premise of the UDC LD service is that it ought to support practical use of the scheme as well as to protect UDC publishing in a way that its future is safeguarded. This specifically means that only a small part of the UDC data shall be published as LOD and most of the UDC LD content would be license protected, i.e., LD "behind the barrier." The UDC LD-based terminological service must support the following features:

1. Programmatic access to:
   a) One LOD set: the UDC Summary containing 3,000 classes (under CC BY-NC-ND 2.0 license); and,
   b) Two LD sets behind a UDC MRF license barrier:
      i) Abridged edition (12,000 classes); and,
      ii) UDC MRF (72,000 classes), including all twenty versions of the UDC MRF and historical data comprising 13,000 cancelled (deprecated) classes and their redirections to new classes;
2. A UDC Look-up service that:
   a) parses and resolves (interprets) a classmark originated from bibliographic data and links its components to relevant records in the RDF data store; and,
   b) upon request supplies URI(s) for UDC classmarks or the full RDF records.

The architecture of the UDC Look-up service has the following components:

1. RDF stores (three Virtuoso databases: the UDC Summary, the Abridged edition, and the UDC MRF) with SPARQL endpoints accessible only via a restricted RESTful API layer which uses pre-designed SPARQL templates for query execution.
2. Apache web server and custom written UDC parser written in PHP and Java. The Authentication process is handled by standard shared and private authentication keys. The HTTP/Get parameters and the HTTP headers inform the server about the type of desired result (e.g., HTML, RDF-Turtle, JSON).

Although the UDC Look-up service is planned primarily as an API for programmatic interaction it will also have an html interface for human interaction with the service. It is assumed that the API would be queried by programs submitting simple or complex UDC classmarks either to get correct URIs for UDC codes or to retrieve complete RDF records. The HTML interface allows humans to verify and explore the provided classmarks in which the parsetree, versions, and RDF translations are expressed. The most important part of this service is the "UDC interpreter", i.e., a program that parses complex UDC strings. This interpreter is based on a series of algorithms developed in an earlier project by Attila Piros (cf. Piros 2017). The UDC notation system allows for 100% accuracy in parsing of UDC strings using several groups of algorithms. Figure 1 shows an HTML interface in which a complex UDC number is split into components that, in this case, are all valid UDC classes. The service executes queries against the UDC Summary, the UDC Abridged edition or the UDC MRF and in the second step it generates an RDF representation of the information selected by the user/machine from the previous step. For clarity, the terms shown in bold underline font in Figure 1 are resolvable primitive UDC terms.

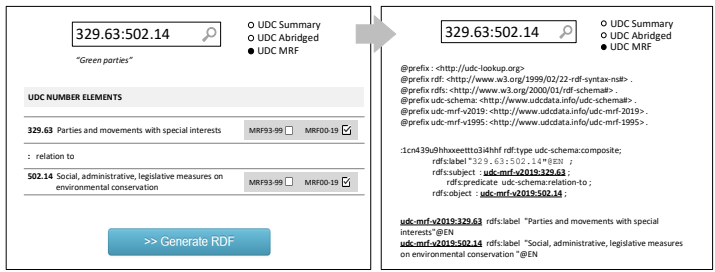

Figure 1. UDC Look-up service and interpreter

## 3.2 Steps in publishing UDC LD
This section outlines some of the key decisions and steps in the UDC LD service design. They broadly follow the ten step guidelines described by Siebes et al. (2019).

### 3.2.1 Selection of data
An important effort in this project was put into the strategic thinking and planning of UDC LD and in particular having to do with the selection of data to be published. The UDC Summary, the UDC Abridged Edition, and the UDC MRF are maintained in different MySQL databases and the same set-up is replicated for the RDF store (three Virtuoso databases). The selection of these three datasets is based on the well-established practice in UDC data use and publishing. They are representative of two kinds of access to UDC data: open access and access through a UDC MRF license requiring an authentication process based on authentication tokens (managed outside the service itself). With respect to the supported languages, the UDC Summary contains language data in 57 languages. However, in this phase the Abridged and MRF datasets are available only in English. UDC data comprise many data elements that are required for data management and publishing, for the LD we selected only 14 data elements. In terms of sequence of data release, the UDC Summary (the LOD set) was given priority due to the large community of users.

### 3.2.2 URI name strategy
The UDC namespace was already established in 2011 and will remain as https://udcdata.info. The UDC experience shows that the decisions regarding the URI are far from being trivial. In the 2011 LD version, we opted for URIs that had the format of the following example: "udcdata.info/068288" in which the number "068288" represented a UDC record identifier for the notation =162.3 Czech language. An important reason for not including, at the time, UDC notation in the URI was the practice of the occasional re-use of deprecated notations (usually after 10 or more years). Thus, notation on its own was considered an unreliable identifier. Once historical versions of the MRF are included in linked data, a version code can be used to contextualise the notation, so we opted for a structured URI that includes UDC notation in the following format: "udcdata.info/MRF93/=162.3." In this example, the element "MRF93" represents the earliest MRF version in which this UDC classmark appeared, i.e., the version in which it was introduced for the first time. The advantage of this approach is that it makes easier for libraries to generate classmark queries to be launched against the UDC Look-up service and allows for human control of URIs (should this be required). An inconvenience with this approach is that UDC classmarks contain symbols and punctuations that are encoded automatically as they get processed, thus udcdata.info/MRF93/=162.3 becomes udcdata.info/MRF93/%3D162. This change of the URI format means that a new service must contain the mapping between the old 2011-2019 URIs and the new URI systems.

### 3.2.3 Use scenarios, serialization and resolution of UDC codes and URIs

When it comes to the Linked Data serialization of the UDC data source, we have to consider various scenarios in which the UDC namespace will be accessed. Since the service is primarily aimed for machine access, we need to have disambiguation mechanisms combined with a clear guidance to make the programmers aware of the various choices that apply. For example, often the only information libraries have about the UDC is the classmarks and the location of the UDC Look-up service. They are not aware of the UDC MRF versions, including whether classmarks contain valid or deprecated numbers or whether they have license, i.e., authentication token, to query full UDC data. Their queries may have the following format "udcdata.info/681.3(035)." The UDC Look-up service will parse and resolve the query indicating that notation 681.3 is deprecated and replaced by 004 and may return an RDF statement with sets of URIs expressing the relationship between these two numbers. If later at time an entity (machine or human) without an access key for this dataset tried to query these URI's at the UDC namespace, the authentication layer would prevent this request from being executed and return a meaningful error message, eventually combined with some sparse information about the result of the query (e.g., a superclass which the concept shares both from the MRF version and the UDC-summary version).

### 3.2.4 Selection of RDF schema

Following the parsing stage, URIs for individual classmark components and their grouping are generated using RDFs. For the full RDF records we use the SKOS format as it is widely used in the KOS publishing community. Equally, we wanted to maintain continuity with the 2011-2019 UDC linked data version. Below we can see the current mapping between UDC MRF data elements[3] and the SKOS schema, which is extended by UDC sub-elements (in italics):

| | | |
|---|---|---|
| UDC number (notation) | skos:notation | |
| class identifier | skos:Concept | |
| broader class | skos:broader | |
| caption | skos:prefLabel | |
| including note | skos:note | *udc:includingNote* |
| application note | skos:note | *udc:applicationNote* |
| scope note | skos:scopeNote | |
| examples | skos:example | |
| see also reference | skos:related | |
| revision history | skos:historyNote | *udc:revisionHistory* |
| introduction date | skos:historyNote | *udc:introductionDate* |
| cancellation date | skos:historyNote | *udc:cancellationDate* |
| replaced by | skos:historyNote | *udc:replacedBy* |
| last revision data | skos:historyNote | *udc:lastrevisionDate* |

In the future, we plan to move towards more formalized schemas from the OWL stack. This would enable a precise formalization that allows semantic verification of classmark strings, the vocabulary itself (e.g., when new concepts with their constraining properties are added or removed in future releases), and rich inference via transitivity, reflexivity, etc.

**4.0 Conclusion and future work**

As in many LD projects, the planning phase of both the UDC and BCC LD projects took more time than originally anticipated. What is often underestimated is that the translation or transference of a resource to another medium or another technology is not merely a technological enterprise but is in essence coupled to a variety of research problems. The process can be compared to the mapping of vocabularies to each other, which is also not a mere mechanical process but entails all kinds of research and editorial decisions, which in turn will influence how a KOS resource is further used. To operate on the scale of the web and with in principle unlimited outreach and spreading, the problem is only augmented. For both the UDC and BCC, key decisions had to be made through the combination of expertise in LD technologies and publishing models, on the one hand, and expertise in the UDC or BCC schemes, datasets, and publishing models, on the other. More time for reflection, research, learning and discussion than envisioned was necessary in all key stages of the project. UDC and BCC are KOSs of a different type. The BCC is newer, experimental and still growing structurally. The UDC is one of the few authoritative KOSs for bibliographic databases, implemented widely, and based on a long history and fully developed KO principles of further development and implementation. Hence, the requirements for the LD publication are very different, and combining them was not part of the DiKG project. In this paper, we describe the different challenges those two KOSs are exposed to during the LD publication.

Planning and developing of the UDC namespace in the form of a Look-up service presented challenges primarily because it is both a new and a complex approach to KOS publishing, also in the realm of established semantic web practices. Web-supported access to the UDC for humans based on a multi-tier license access that combines free access to part of the resource with licensed access for experts. This needs to be mimicked in the LOD transition. In our approach, LOD and "LD behind the license barrier" models of publishing are combined and involve three different levels of classification data aimed at different audience and use scenarios. An important part of the UDC LD cloud is its historical data that will hopefully enhance the usability of UDC in the bibliographic domain where historical and obsolete classification data appear frequently. The most novel and key function to the Look-up service is the UDC interpreter. The UDC namespace is envisaged as a one-stop shop for querying and validating UDC data and it also illustrates a more complex, but hopefully more robust, model of KOS publishing as linked data. This UDC namespace offers a good environment for linked data and library linked data study and research on KOS alignments and integration.

**Notes**
1. Digging Into the Knowledge Graph (DIKG). https://diggingintodata.org/awards/2016/project/digging-knowledge-graph
2. Basic Concepts Classification. https://sites.google.com/a/ualberta.ca/rick-szostak/research/basic-concepts-classification-web-version-2013
3. The UDC MRF data elements schema is available at: http://www.udcc.org/files/udc_data_elements_mrf11.pdf

**References**

Eito-Brun, Ricardo. 2018. "The Role of Knowledge Organization Tools in Open Innovation Platforms." In *Challenges and Opportunities for Knowledge Organization in the Digital Age: Proceedings of the Fifteenth International ISKO Conference, 9-11 July 2018, Porto, Portugal*, ed. Fernanda Ribeiro and Maria Elisa Cerveira. Advances in Knowledge Organization 16. Baden-Baden: Ergon, 666-73.

Marcondes, Carlos. H. 2018. "Culturally Relevant Relationships: Publishing and Connecting Digital Objects in Collections of Archives, Libraries, and Museums over the Web." In *Challenges and Opportunities for Knowledge Organization in the Digital Age: Proceedings of the Fifteenth International ISKO Conference, 9-11 July 2018, Porto, Portugal*, ed. Fernanda Ribeiro and Maria Elisa Cerveira. Advances in Knowledge Organization 16. Baden-Baden: Ergon, 539-48.

Martínez-Ávila, Daniel, Richard P. Smiraglia, Rick Szostak, Andrea Scharnhorst, Wouter Beek, Ronald Siebes, Laura Ridenour and Vanessa Schlais. 2018. "Classifying the LOD Cloud: Digging into the Knowledge Graph." *Brazilian Journal of Information Studies: Research Trends* 12, no. 4: 6-10.

Piros, Attila. 2017. "The Thought Behind the Symbol: About the Automatic Interpretation and Representation of UDC Numbers". In *Faceted Classification Today: Theory, Technology and End Users: Proceedings of the International UDC Seminar 2017, London (UK), 14-15 September*, ed. Aida Slavic and Claudio Gnoli. Würzburg: Ergon Verlag, 203-18.

Siebes, Ronald, Gerard Coen, Kathleen Gregory, and Andrea Scharnhorst. 2019. "Linked Open Data. 10 Things toward the LOD Realm: The "I" in FAIR in a Semantic Way." Zenodo https://doi.org/10.5281/zenodo.3471806

Slavic, Aida. 2017. "Klasifikacija i Library Linked Data (LLD) = Classification and Library Linked Data (LLD)". In *Predmetna Obrada: Pogled Unaprijed: Zbornik Radova*, ed. B. Purgaric and S. Spiranc. Zagreb: HKD, 13-37.

Szostak, Rick, Andrea Scharnhorst, Wouter Beek and Richard P. Smiraglia. 2018. "Connecting KOSs and the LOD Cloud." In *Challenges and Opportunities for Knowledge Organization in the Digital Age: Proceedings of the Fifteenth International ISKO Conference, 9-11 July 2018, Porto, Portugal*, ed. Fernanda Ribeiro and Maria Elisa Cerveira. Advances in Knowledge Organization 16. Baden-Baden: Ergon, 521-29.

Szostak, Rick. 2019. "The Basic Concepts Classification." In *ISKO Encyclopedia of Knowledge Organization*, edited by Birger Hjørland and Claudio Gnoli. https://www.isko.org/cyclo/bcc